\documentclass[aps, superscriptaddress, twocolumn, nofootinbib]{revtex4}
\usepackage{amsmath, amssymb, mathtools, esint}
\usepackage[utf8]{inputenc} 
\usepackage{graphicx, fancybox}
\usepackage{epstopdf}
\usepackage{color}
\usepackage{cancel}
\usepackage{listings}
\usepackage{rotating}
\usepackage{hyperref}
\usepackage{filecontents}
\usepackage{siunitx}
\usepackage{dcolumn}

\newcolumntype{e}[1]{D{.}{.}{#1}}
\newcommand\mc[1]{\multicolumn{1}{c|}{#1}}

\newcommand{\abs}[1]{\left| #1 \right|}

\newcommand{\er}{\;\hat r}
\newcommand{\et}{\;\hat \theta}
\newcommand{\ep}{\;\hat \phi}

\newcommand{\ex}{\;\hat x}
\newcommand{\ey}{\;\hat y}
\newcommand{\ez}{\;\hat z}

\newcommand{\vn}{\;\vec \nabla}
\newcommand{\vb}{\;\vec B}
\newcommand{\vm}{\;\vec \mu}

\newcommand{\ct}{\cos \theta}
\newcommand{\st}{\sin \theta}
\newcommand{\cp}{\cos \phi}
\renewcommand{\sp}{\sin \phi}

\begin{document}
\title{Nonlinear Semi-Classical 3D Quantum Spin}
\author{J.J. Heiner}
\email[]{jheiner2@uwyo.edu}
\affiliation{Department of Physics and Astronomy, University of Wyoming}
\author{J.D. Bodyfelt}
\email[]{jdbodyfelt@gmail.com}
\affiliation{Centre for Theoretical Chemistry and Physics, Massey University}
\author{D.R. Thayer}
\email[]{drthayer@uwyo.edu}
\affiliation{Department of Physics and Astronomy, University of Wyoming}
\begin{abstract}
In an effort to provide an alternative method to represent a quantum spin, a precise 3D nonlinear dynamics method is used.  A two-sided torque function is created to mimic the unique behavior of the quantum spin.  A full 3D representation of the magnetic field of a Stern-Gerlach device was used as in the original experiment.  Furthermore, the temporarily driven nonlinear damped model exhibits chaos, but stuggles to be consistent through azimuthal angles in reproducing the quantum spin statistics.
\end{abstract}
\maketitle
\section{Introduction}
Scientists have questioned how quantum spins evolve into one of two states \cite{Feynman, Bellac, Platt}.  In a recent publication it was discussed that it may be possible to understand the quantum mechanical spin state evolution, or quantum mechanical wave collapse, in a similar method used in deterministic chaos, which does not violate the Bell inequalities \cite{Thayer_Jafari, Thayer}.  In follow up on that suggestion, a 2D nonlinear semi-classical perturbation model was developed and the results relatively produced the correct statistical quantum expectations \cite{2D_qspin}.  This model was limited to a magnetic field from a current loop, but here the model is expanded into 3D.  Furthermore, the exact 3D magnetic field from a Stern-Gerlach device is calculated and used in this research publication.

\begin{figure}
\includegraphics[scale=0.35, trim={0 7cm 0 5cm}]{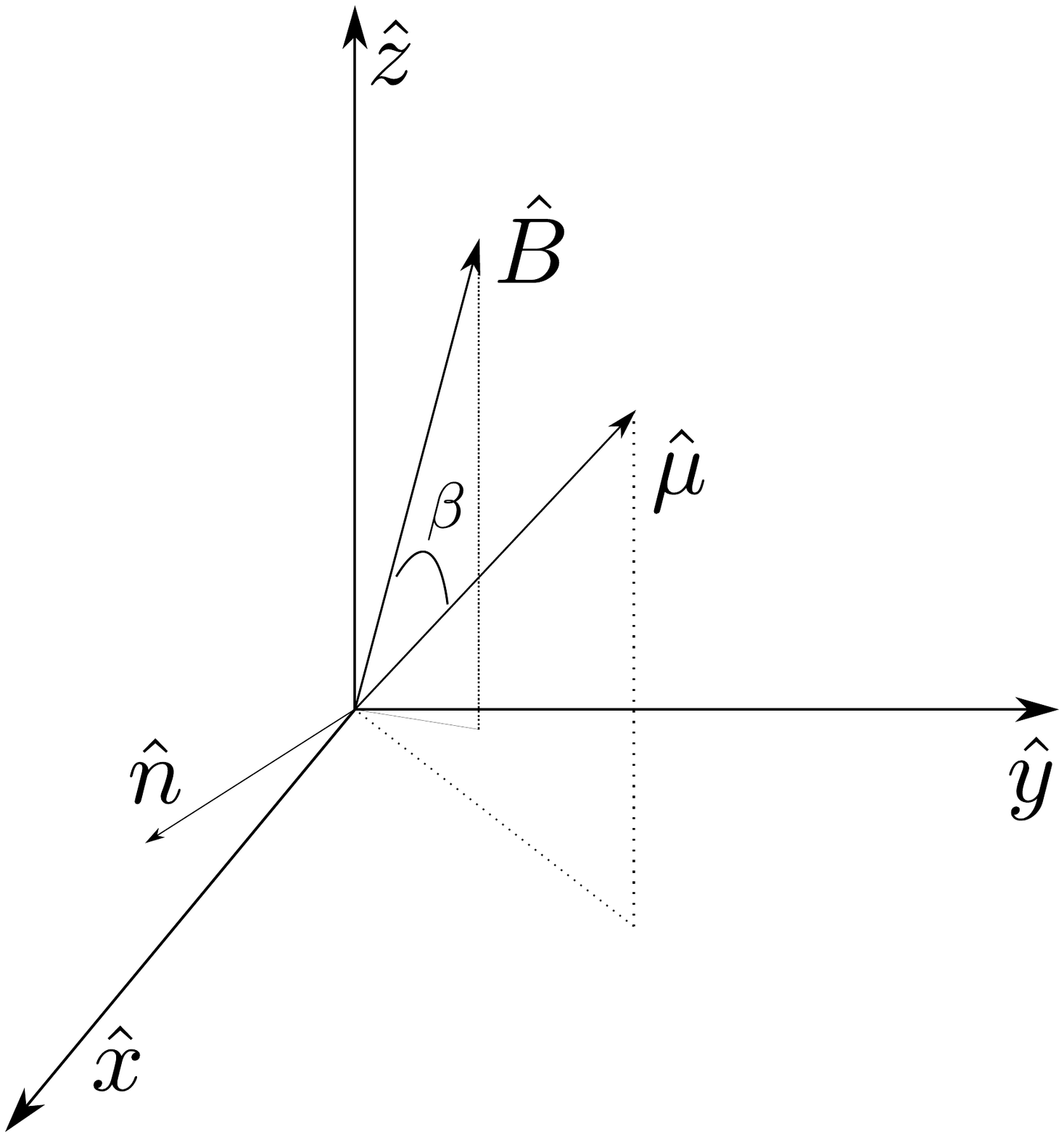}
\caption{\label{fig:3D_Geometry_Spin}3D Geometry for the Semi-Classical Spin Model in the presence of a magnetic field, $\vec B$, along with the normalized torque $\hat n$}
\end{figure}

The geometry used to describe the relationship between the unit quantum spin, $\hat \mu$, and the unit magnetic field, $\hat B$, (which is rapidly evolving with respect to the quantum spin) can be seen in figure \ref{fig:3D_Geometry_Spin}, where the angle of separation is $\beta$, and the unit vector of magnetic torque rotation is represented as $\hat n$.

The probability that the quantum spin will collapse in the direction of the magnetic field, spin up, and the probability that it will collapse in the opposite direction, spin down, is given as

\begin{equation}
\begin{split}
P_\uparrow &= \cos^2 \beta/2, \\ P_\downarrow &= \sin^2 \beta/2.
\end{split}
\label{eq:Percentage}
\end{equation}

In the presence of a nonuniform magnetic field, once the spin has collapsed into the spin up or down state there will be a classical force that acts on the spin magnetic moment.  The force is written as

\begin{equation}
F=\vn (\vm \cdot \vb),
\end{equation}

\noindent where often the assumption inside the Stern-Gerlach device is

\begin{equation}
F \simeq \mu_z \frac{\partial B_z}{\partial z} \hat z,
\label{eq:force_simp}
\end{equation}

\noindent and the force value can be either positive or negative depending on the direction of $\mu_z$ \cite{Stern}.  
\section{Semi-Classical Torque: Moment Dynamics}
As the spin magnetic moment, $\vec \mu$, is not a classical magnetic moment, it is necessary to consider the very peculiar aspect as there appears to exist two stable equilibrium locations.  However, the classical magnetic moment torque has two equilibrium locations which depend on the angle, $\beta$, one being stable at $\beta = 0$ and the other being unstable at $\beta = \pi$.  This can be easily observed in the torque of a classical dipole moment in a magnetic field:

\begin{equation}
\vec \tau_\textrm{c} = \vm_c \times \vb = \mu_c B \sin (\beta) \, \hat n,
\label{eq:torque_classical}
\end{equation}

\noindent where

\begin{equation}
\hat n = \frac{\vm \times \vb}{\lvert \vm \times \vb \rvert}.
\end{equation}

\begin{figure}
\includegraphics[scale=.6, trim={0 0cm 0 0cm}]{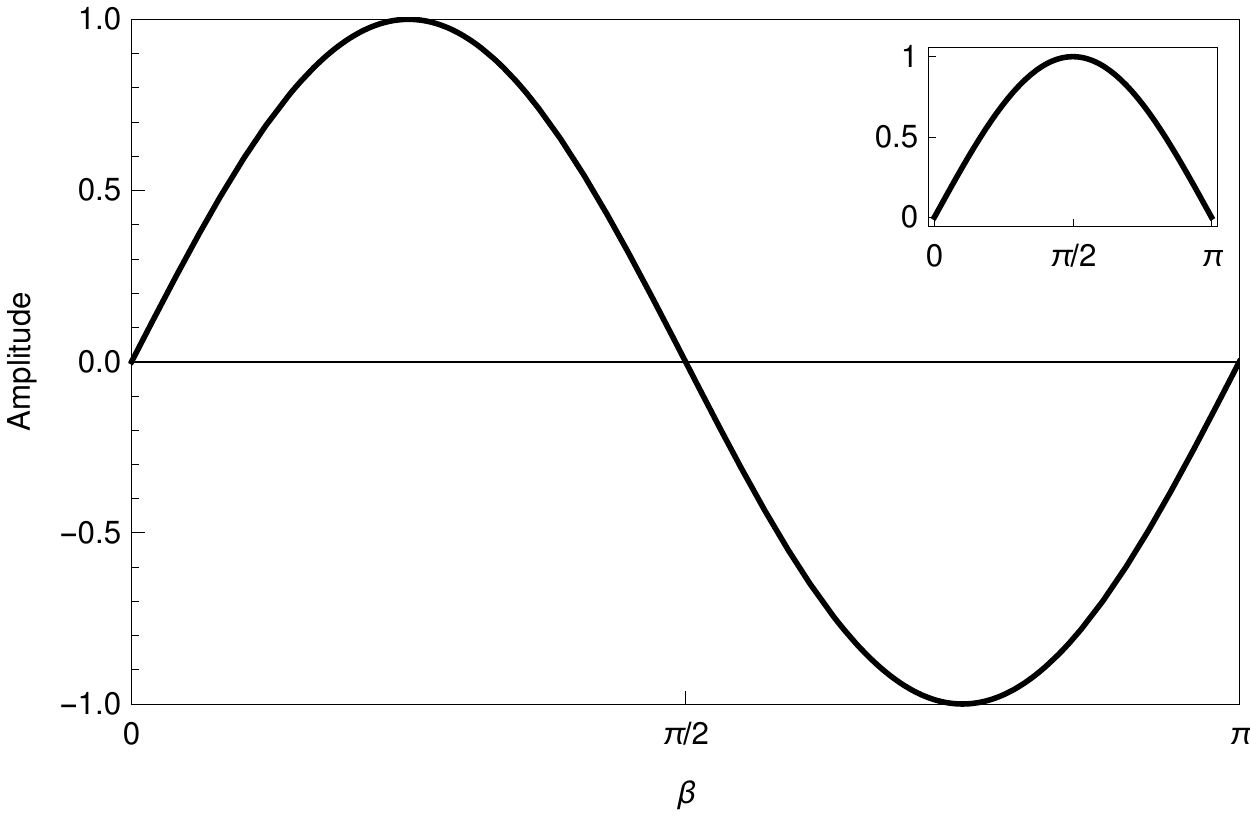}
\caption{\label{fig:semi_class_torque}The normalized semi-classical spin torque is shown from equation \ref{eq:torque_semi_classical} with $c = 2$.  The classical torque function, as a reminder, is in the inset.}
\end{figure}

The sinusoidal function that arises in equation \ref{eq:torque_classical} can be modified into a semi-classical torque representation so there are two stable equilibriums at $\beta = 0$ and $\beta = \pi$ (representing the final evolution state of the quantum spin).  A function that fills this unique semi-classical torque behavior is

\begin{equation}
\vec \tau_\textrm{sc} = -\mu B \sin (\beta) \, \tanh (c \, (\beta -\pi/2 ) ) \, \, \hat n,
\label{eq:torque_semi_classical}
\end{equation}

\noindent where $c$ is a parameter to change the sharpness of the hyperbolic tangent function (the negative arises due to the hyperbolic tangent function).  The comparison of this new semi-classical torque to the classical torque can be seen in figure \ref{fig:semi_class_torque}, as normalized figures.

\begin{figure}
\includegraphics[scale=.7, trim={.4cm 0.5cm 0.4cm 3.2cm},clip]{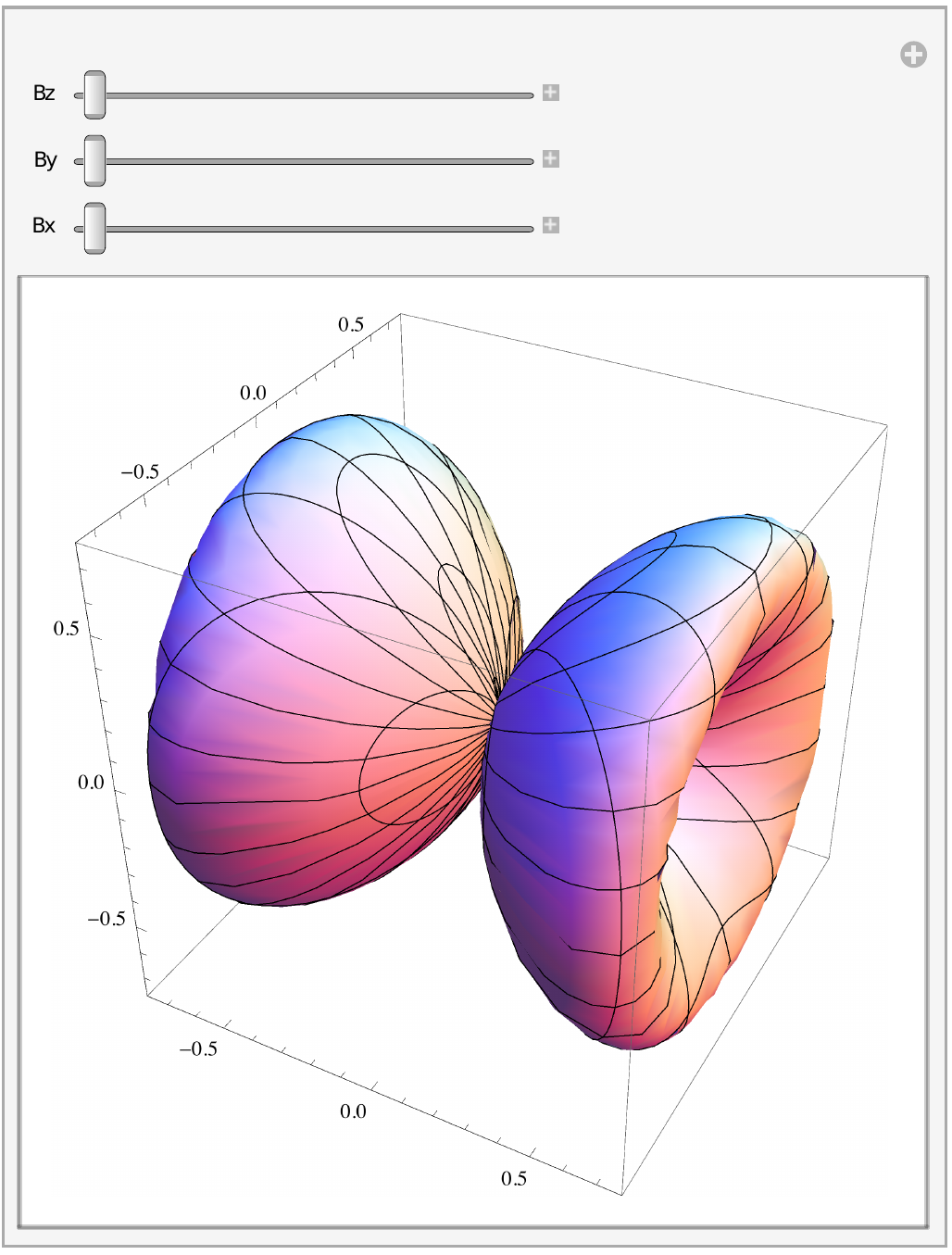}
\caption{\label{fig:torque_3D}A normalized 3D semi-classical dual torque of a quantum spin subjected to a magnetic field pointing along the axis of highest symmetry.  As further explained in supplementary \ref{sec:EOM}, the torque magnitude model can easily be changed and adapted.}
\end{figure}

It is important to conclude that many torque models fit the qualifications of two stable equilibria.  For example figure \ref{fig:torque_3D} shows a normalized 3D plot surface, which is symmetric about the magnetic field, $\vec B$, and follows it as it dynamically evolves.

\begin{figure}
\includegraphics[clip, trim={2cm 14cm 3cm 1cm}, scale=0.5]{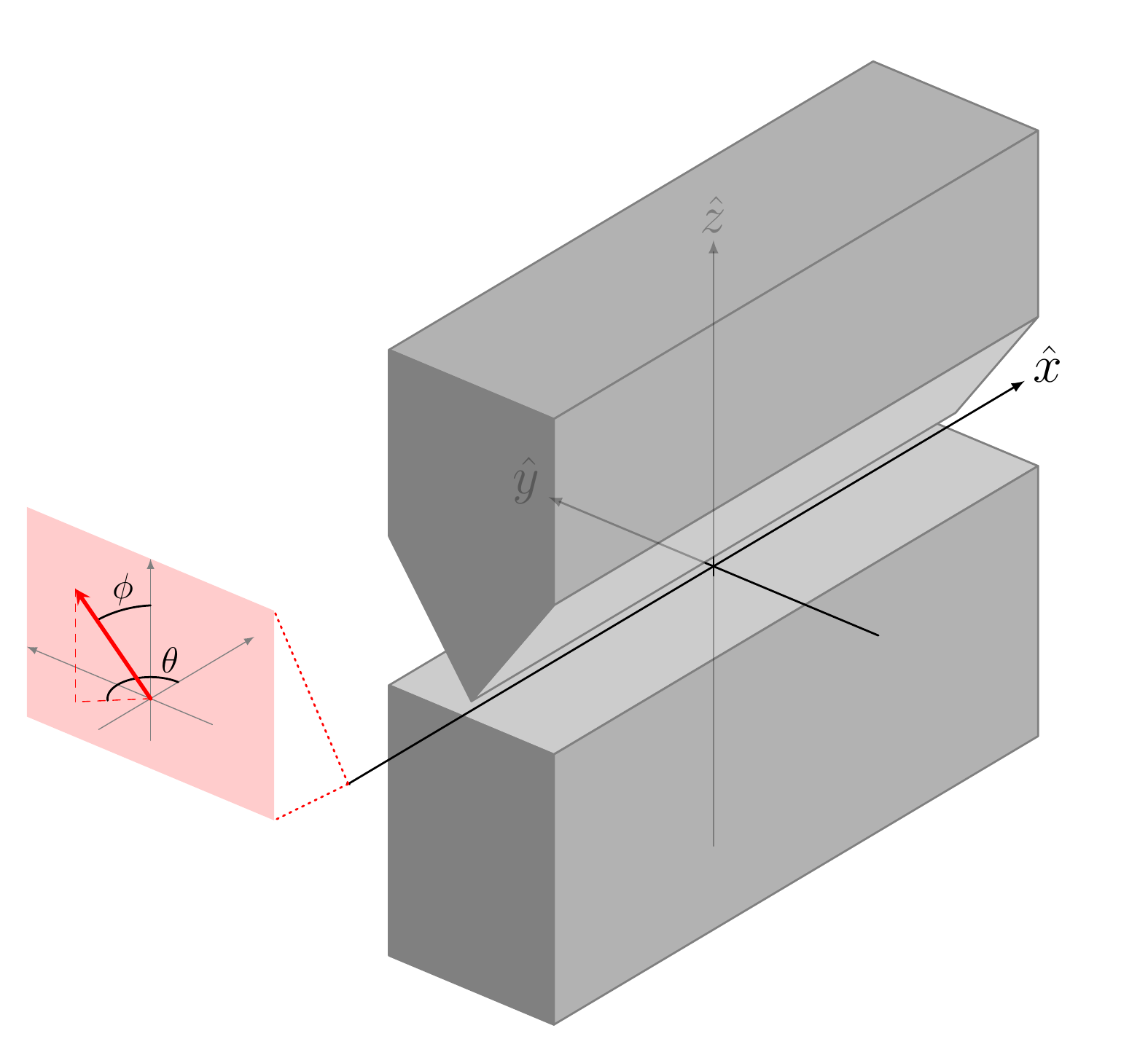}
\caption{\label{fig:setup_figure}The overall view of the Stern-Gerlach device and the definition of the coordinate system being used by the magnetic field and the quantum spin moment.}
\end{figure}

By treating the magnetic moment, $\vec \mu$, as a rod, $\vec r$, an evolution in time under spherical coordinates where $\phi$ is the angle off of the $z$ axis and $\theta$ is the azimuthal angle around the $z$ axis, starting at the $x$ axis as in figure \ref{fig:setup_figure}, the angular velocity for the spin moment in cartesian coordinates is

\begin{equation}
\vec \omega = \frac{\vec r \times \vec v}{r^2}.
\end{equation}

Differentiating angular velocity gives angular acceleration:

\begin{equation}
\vec \alpha = \frac{d \vec \omega}{dt} =
\cancelto{0}{\frac{\vec v \times \vec v}{r^2}} + \frac{\vec r \times \vec a}{r^2} - 2 \dot r \frac{\vec r \times \vec v}{r^3},
\end{equation}

\noindent which in terms of our coordinate system can be written\footnote{For the full derivation see section \ref{sec:ang_acc} of the supplementary information} as

\begin{eqnarray}
\vec \alpha &=& 
\begin{pmatrix}
    \ddot \phi \st - {\dot \theta}^2 \st \sp \cp + \\ 2\dot \theta \dot \phi \ct \cos^2 \phi + 
    \ddot \theta \ct \sp \cp \\[6pt]
    -\ddot \phi \ct + {\dot \theta}^2 \ct \sp \cp + \\ 2\dot \theta \dot \phi \st \cos^2 \phi +
     \ddot \theta \st \sp \cp \\[6pt]
     -2\dot \theta \dot \phi \cp \sp - \ddot \theta \sin^2 \phi
    \end{pmatrix}.
\end{eqnarray}

With the angular acceleration being in cartesian coordinates, it can be related to the torque as

\begin{equation}
\boldsymbol{\hat I} \vec \alpha = \vec \tau,
\label{eq:EoM_torque}
\end{equation}

\noindent where the torque term is the sum of all torques.  Included in the torque terms is a linear angular dissipation force, i.e. $\vec{\tau}_{diss}=b \, \vec \omega$, where

\begin{eqnarray}
\vec \tau &=& 
\begin{pmatrix}
    -\mu B \sin (\beta) \, \tanh (c \, (\beta -\pi/2 ) ) \, \, \hat n \cdot \hat x \\ -b(\dot \phi \st + \dot \theta \sp \cp \ct) \\[6pt]
    -\mu B \sin (\beta) \, \tanh (c \, (\beta -\pi/2 ) ) \, \, \hat n \cdot \hat y \\ -b(-\dot \phi \ct + \dot \theta \sp \cp \st) \\[6pt]
     -\mu B \sin (\beta) \, \tanh (c \, (\beta -\pi/2 ) ) \, \, \hat n \cdot \hat z \\ -b(-\dot \theta \sin^2 \phi)
    \end{pmatrix},
\end{eqnarray}

\noindent and $b$ is a dissipation factor.

Although a threshold criterion has been presented in terms of a moment of inertia, $\boldsymbol{\hat I}$, that roughly separates quantum behavior from classical \cite{inertia_threshold}, this research publication proposes and is not the first to use a moment of inertia tensor for a quantum spin moment \cite{Kikuchi}.  To first order approximation, the moment of inertia tensor for a quantum spin should be a thin rod, which the reader is reminded:

\begin{eqnarray}
\boldsymbol{ \hat I} &=& I 
\begin{bmatrix}
1 & 0 & 0 \\
0 & 1 & 0 \\
0 & 0 & 0
\end{bmatrix},
\end{eqnarray}

\noindent where the scaler $I$ along with $b$ will be used as adjustable parameters in the simulation.

To obtain the two acceleration equations of motion\footnote{See section \ref{sec:EOM} of the supplementary information for equations of motion for $\ddot \theta$ and $\ddot \phi$} for the quantum spin, namely $\ddot \theta$ and $\ddot \phi$, linear algebra is used to solve equation \ref{eq:EoM_torque}. 
\section{3D Magnetic Field of the SGD}
The force on a quantum spin is due to a magnetic field; therefore, the magnetic field of the Stern-Gerlach device will be discussed first.  Since a full representation of the magnetic field was needed, and there exist a magnetic field prior to the spin entering the Stern-Gerlach device, the full magnetic field is calculated.  Although some have calulated a 2D magnetic field using a finite element method \cite{Gersem}, the complete 3D magnetic field was obtained by the Biot-Savart law:

\begin{equation}
B(r)=\frac{\mu_0}{4\pi} \int \frac{J(r') \times (\vec r - \vec {r'})}{\abs{\vec r - \vec {r'}}^3} \, d\tau',
\end{equation}

\noindent where $\vec{r'}$ is the vector from the origin to the source point and $\vec{r}$ is the vector from the origin to the field point.  $J(r')$ is the current density written also in terms of the magnetization, $\textrm{M}$, as a bound volume current, $\textrm{J}_b = \nabla \times \textrm{M}$, plus the bound surface current, $\textrm{K}_b = \textrm{M} \times \textrm{\^{n}}$, where \^{n} is the normal to the surface unit vector.

\begin{figure}
\includegraphics[clip, trim={0cm 0cm 0cm 0cm}, scale=.65]{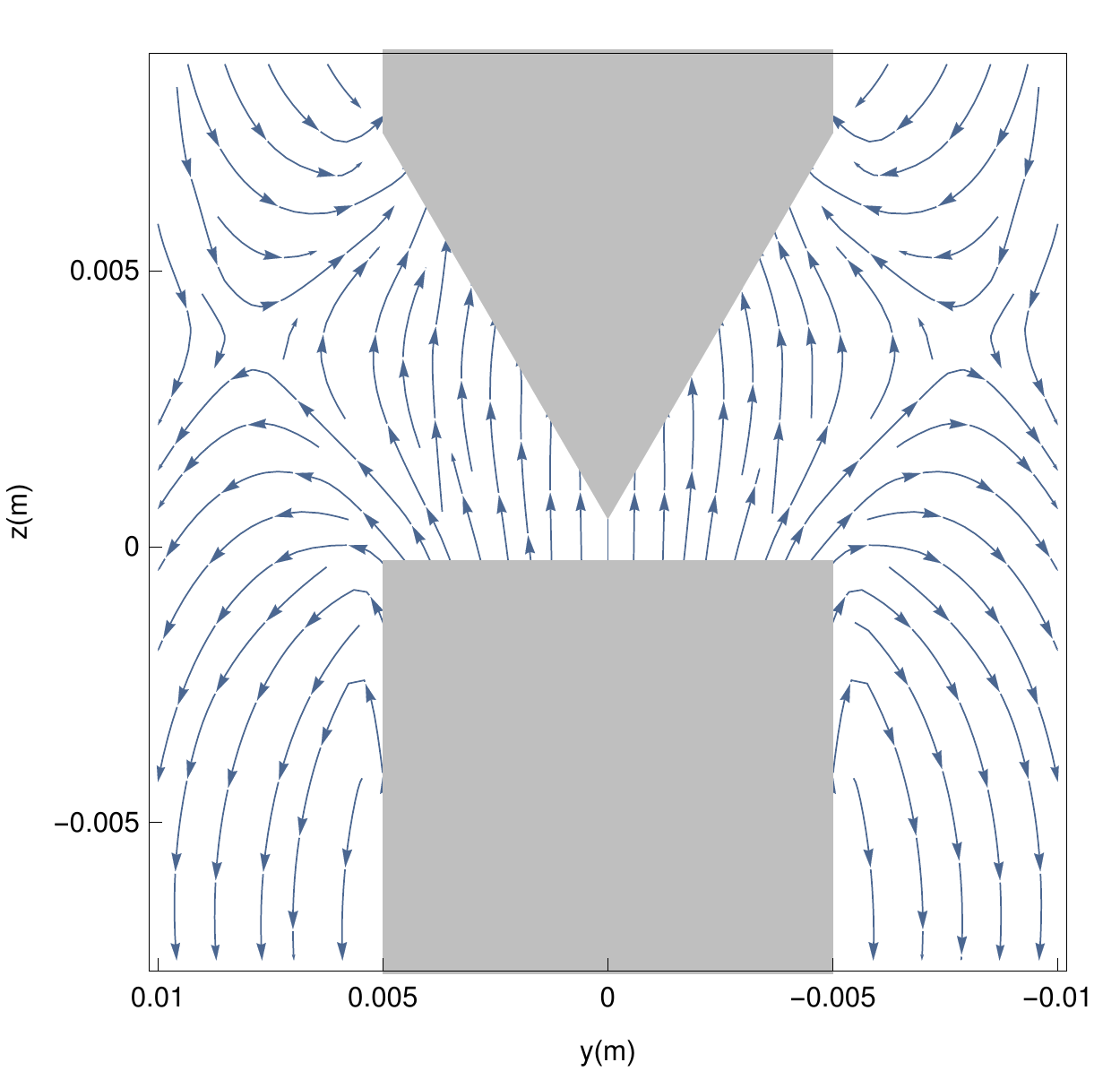}
\caption{\label{fig:SGD_B_field}This stream slice view of the Stern-Gerlach device is headon in the $y$-$z$ plane at $x=0$.  The magnetic field was obtained from the Biot-Savart law.}
\end{figure}

An analytical solution for the magnetic field of the Stern-Gerlach device, modeled after figure \ref{fig:setup_figure}, was obtained using mathematica\footnote{The dimension definitions used in the Stern-Gerlach device can be seen in the supplementary information section \ref{sec:SGD}}.  A stream slice of the magnetic field can be seen in figure \ref{fig:SGD_B_field}.  Since an analytical solution was obtained, it was easy to also obtain analytical solutions for the curl and the divergence of the magnetic field to be used in the force kinematics.

\section{Force: Carrier Kinematics}
\begin{figure}
\includegraphics[clip, trim={0cm 0cm 0cm 0cm}, scale=.75]{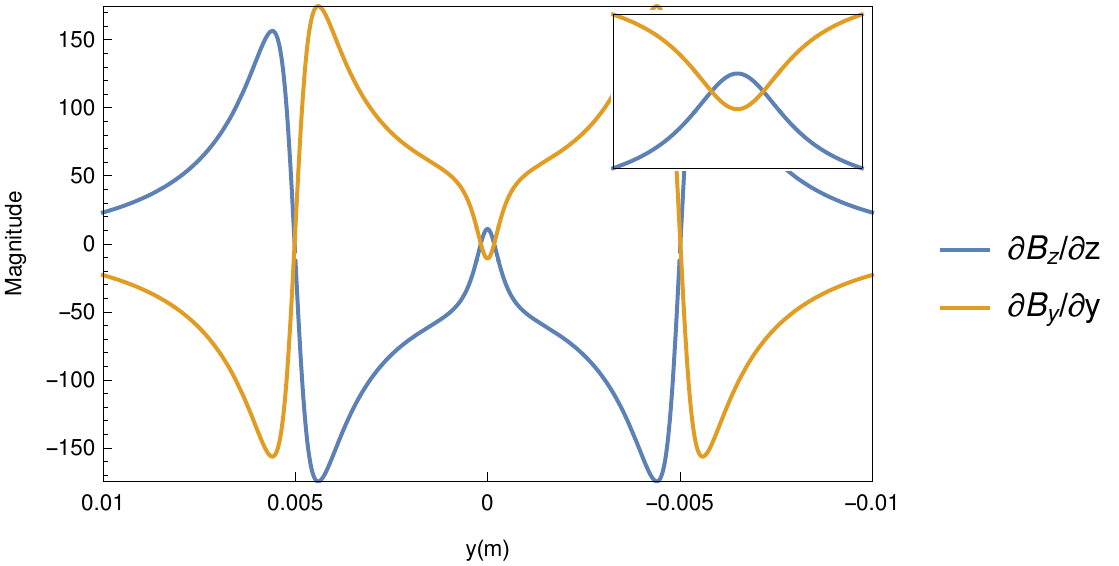}
\caption{\label{fig:divergence_B}The divergence of the magnetic field of the Stern-Gerlach device along the $y$ axis at $x=0$ and $z=0$.  The values from $\partial B_x / \partial x$ are not included as it was relatively zero since $\vec \nabla \cdot \vec B = 0$.  The inset is the blown up shot of the divergence around $y=0$.}
\end{figure}

Many studies acknowledge that there also exists a gradient in the $x$ and $y$ directions since $\vec \nabla \cdot \vec B = 0$ \cite{SG_div_B_Naval, SG_div_B_BYU, Cruz, Lieberman, Stenson, Aharonov}.  Although

\begin{equation}
\frac{\partial B_z}{\partial z} \cong -\frac{\partial B_y}{\partial y},
\end{equation}

\noindent inside the Stern-Gerlach, since  $\mu_y$ averages to zero this force term will also average to zero \cite{Alstrom}(the magnitudes of the divergence can be seen in figure \ref{fig:divergence_B}).  However, few analyze and fail to mention the full force which includes curling terms on the same order of magnitude as the divergence terms \cite{Singh}:

\begin{equation}
\begin{split}
F=&\vn (\vm \cdot \vb),\\
=& \,(\vm \cdot \vn)\vb + (\vb \cdot \vn)\vm + \vm \times (\vn \times \vb)\\ &+ \vb \times (\vn \times \vm),\\
=& \,(\vm \cdot \vn)\vb + \vm \times (\vn \times \vb).\\
\end{split}
\label{eq:force}
\end{equation}

The divergence term cooresponds to a traditional lateral force, $\pm \hat z$, whereas the curling terms lead to a drifting force, $\pm \hat y$.  Therefore, a more acurate assumption that helps account for the drifting seen in the actual Stern-Gerlach experiment is

\begin{equation}
F \cong \mu_z \frac{\partial B_z}{\partial z} \hat z + \mu_z \frac{\partial B_z}{\partial y} \hat y -\mu_z \frac{\partial B_y}{\partial z} \hat y.
\label{eq:force_approximation}
\end{equation}

It is important to note that although equation \ref{eq:force_approximation} shows a more accurate assumption of the force on the spin moment, the full force term, equation \ref{eq:force}, will be used unless otherwise stated.

It is important to discuss the divergence of the magnetic field in figure \ref{fig:divergence_B}.  In looking closer at the inset, the divergence force goes from positive to negative.  In the negative region, a spin that is pointing up will now feel a force that is negative.  This is a mathematical treatment irrelevant of the semi-classical torque model that is being presented.  As this is an attribute of the magnetic field of the Stern-Gerlach device, even the traditional wave quantum mechanics would arrive at the same conclusion.  Since there is a magnetic threshold magnitude, under which a spin magnetic moment will not collapse into a state due to time restraints, this is modeled by reducing the force in the negative region:

\begin{equation}
\frac{\partial B_z}{\partial z} \rightarrow \frac{\partial B_z}{\partial z} \, e^ {-(y/\sigma_y)^2},
\label{eq:divergence_approximation}
\end{equation}

\noindent where $\sigma_y$ is approximately the width of the positive divergence.

\section{Quantum Spin Opposing Magnetic Flux}

This research project was interested in describing as many variables as possible to describe the proposed quantum spin model.  As a result, it is often stated as a universal law \cite{Griffiths, Protheroe, Pallavicini} that nature anhors a change in flux and therefore attempts to counter the change in flux.  Since every particle containing a quantum spin has a finite size, that particle will experience a change in flux while exposed to a changing magnetic field.

A search through literature failed to unveil any type of research or prediction into the dynamics that must exist when a quantum spin is opposing the magnetic flux through it's finite size.

Knowing the change in magnetic flux due to the Stern-Gerlach device where $t_\textrm{enter}$ is the time to get from initially outside, $i$, to the final max field, $f$, 

\begin{equation}
\frac{d\Phi_\textrm{SG}}{dt} = \frac{A (B_{\textrm{SG}_f}-B_{\textrm{SG}_i})}{t_\textrm{enter}} \cong \frac{A (B_{\textrm{SG}_f})}{t_\textrm{enter}},
\end{equation}

\noindent and the change in magnetic flux from the quantum spin,

\begin{equation}
\frac{d\Phi_\textrm{QS}}{dt} = \frac{A (B_{\textrm{QS}_f}-B_{\textrm{QS}_i})}{t_\textrm{enter}} \approx \frac{A (B_{\textrm{QS}_f})}{t_\textrm{enter}},
\end{equation}

\noindent a decision can be made whether to include dynamics from the quantum spin opposing/correcting the magnetic flux from the Stern-Gerlach device:

\begin{equation}
\frac{d\Phi_\textrm{SG}}{dt} \overset{?}{\gtrapprox} \frac{d\Phi_\textrm{QS}}{dt},
\end{equation}

\noindent or

\begin{equation}
B_{\textrm{SG}_f} \overset{?}{\gtrapprox} B_{\textrm{QS}_f}.
\end{equation}

The magnetic field from the quantum spin can be approximated by using the residual magnetic field, $B_r$, from a dipole moment:

\begin{equation}
\vec \mu=\frac{1}{\mu_0} \vec B_r V
\end{equation}

\noindent where V is the quantum spin moment volume.  Therefore,

\begin{equation}
B_{\textrm{SG}_f} \lll \frac{\mu \mu_0}{V}
\end{equation}

\noindent by approximately 14 orders of magnitude\footnote{Letting the magnetic field equal $\sim$1T and the radius $r\sim 10^{-15}$m; $1 \lll \frac{10^{-24}10^{-7}}{(10^{15})^3}$.  (It should be noted that the upper limit was taken for the radius.  Had the lower limit been taken the approximation would have been even higher.) }.

Seeing that the mangetic field from the quantum spin can overpower the mangetic field due to the Stern-Gerlach device, the dynamics for such an effect was logically ignored in this research.
\section{Driven-Damped Pendulum}
Although the similarities between the driven-damped pendulum and the 2D semi-classical spin model have been recently discussed \cite{2D_qspin}, it is important that it be discussed here as well.

The equation of motion for a classical driven-damped pendulum, where $\theta'$ is the angle between the mass vector and gravity, is

\begin{equation}
\ddot \theta'=-a'\dot \theta'-b'\sin(\theta')+c'F(t),
\label{eq:DDP}
\end{equation}

\noindent where the constants $a'$, $b'$, and $c'$ are well known constants and $F(t)$ is a driving force \cite{Thornton}.

One representation for the equation of motion for the 3D semi-classical spin model, in it's simplest form is

\begin{equation}
\begin{split}
\ddot \beta \cong &-b \, f(\dot \theta, \dot \phi, \theta, \phi) - g\,f'(\dot \theta, \dot \phi, \theta, \phi) B_z(x,y,z)\\ &+ g\,f''(\dot \theta, \dot \phi, \theta, \phi)B_x(x,y,z),
\end{split}
\label{eq:DDQS}
\end{equation}

\noindent where $g$ is a constant and the primes indicate a different function.  Also, the force due to $B_y$ was neglected in this comparision due to it being relatively small at $y=0$.

Chaos for a pendulum, equation \ref{eq:DDP}, can only occur when the driving force is stronger than the gravity force, i.e. $c'F(t)/b>1$ \cite{Markus, Holmes, Humieres, Kerr}.  Similarly the semi-classical spin model can only be chaotic when $B_x/B_z>1$.  Therefore, $B_x$ serves the same purpose as $F(t)$ in that it is a driving force.

\begin{figure}
\includegraphics[clip, trim={0cm 0cm 0cm 0cm}, scale=.65]{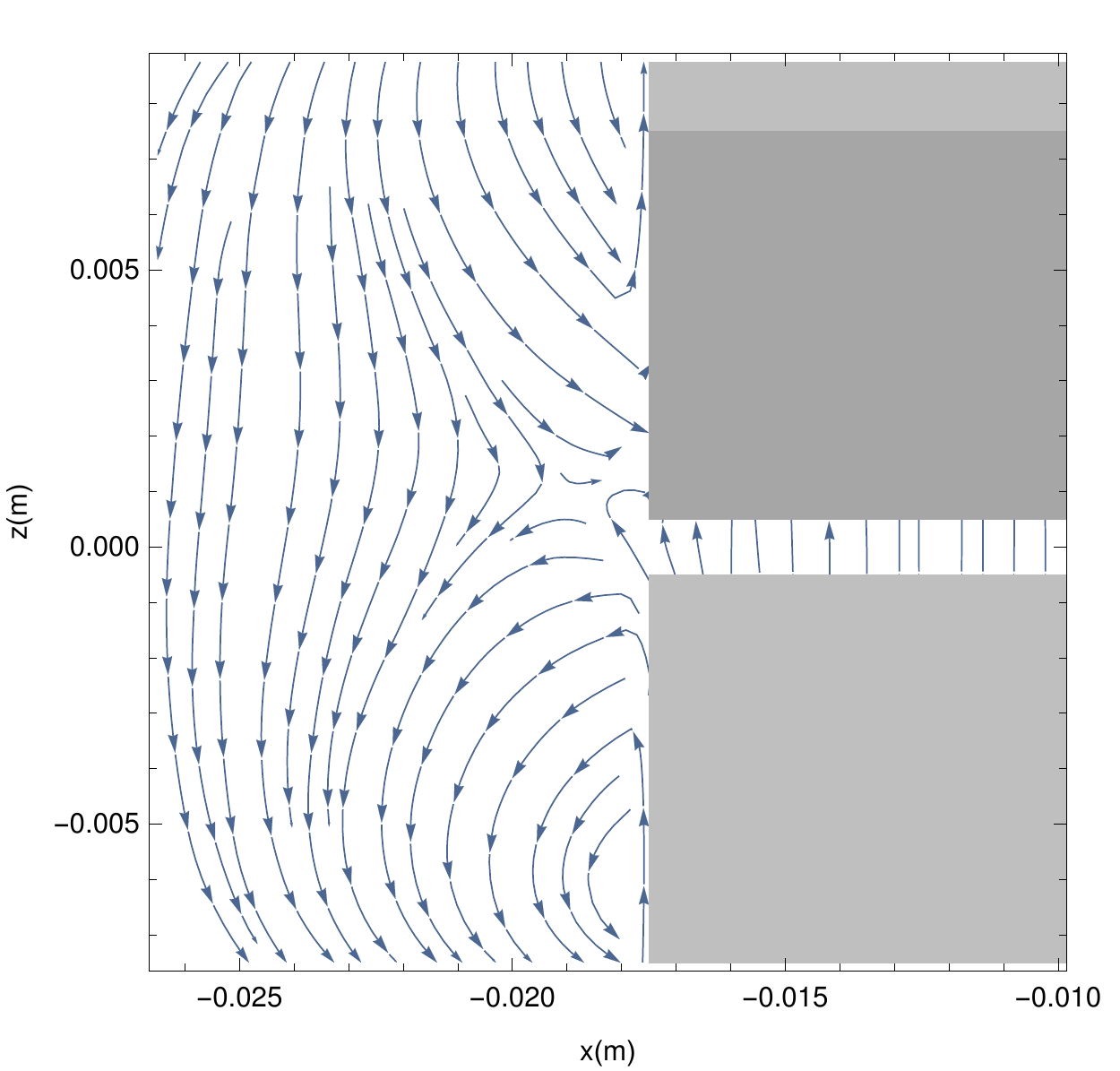}
\caption{\label{fig:SGD_B_field_side}This view of the Stern-Gerlach device is from the side in the $x$-$z$ plane at $y=0$.  The shaded portion indicates the tip of the Stern-Gerlach device.}
\end{figure}

The simulation for a quantum spin begins outside the Stern-Gerlach device and travels in the $\hat x$ direction, figure \ref{fig:setup_figure}.  As seen in figure \ref{fig:SGD_B_field_side}, the magnetic field is dominatly in the $-\hat z$ direction.  Then as the spin approaches the Stern-Gerlach device the magneitc field is dominatly in the $-\hat x$ direction.  It is during this small area of space where $B_x/B_z\gg1$ that the spin will be exposed to only a driving force.  This area is what causes chaos to occur and the quantum statistics to be acheived, i.e. the peculiar idea that a spin which is pointing mostly up has a probability of flipping down as seen from equation \ref{eq:Percentage}.  Furthermore, since $\vec \nabla \cdot \vec B = 0$, there will always exist a Stern-Gerlach-like device where an entering spin will be exposed to a dominately perpendicular magnetic field.
\section{Code}
The numberical method used to step through the resulting equations of motion from equation \ref{eq:EoM_torque}, as fully shown in equation \ref{eq:EOM_full}, that describes the moment dynamics is the forth order Runge-Kutta method.  Since the moment dynamics, which is equatted to a quantum effect, happen at small time scales, only a second order Runge-Kutta method is used to describe the carrier kinematics, equation \ref{eq:force}.

Furthermore, since the moment dynamics is a quantum effect, the forces on the quantum spin were zero until the spin entered the Stern-Gerlach device.  The simulation terminates at the end of the device.  To show the classical trajectory split in the $y$-$z$ plane some distance $d$ away, an elementary physics approach is taken once outside the Stern-Gerlach device:

\begin{equation}
\Delta z = v_z (d/v_x),
\end{equation}   

\noindent where a similar equation can be written for $\Delta y$.

The initial velocity in the $\hat x$ direction is randomly assigned from a gaussian distribution centered at 550m/s, which is very similar to the original Stern-Gerlach experiment.

In solving for $\ddot \theta$, there is a $\sp$ in the denominator.  Therefore as $\phi \rightarrow 0,\pi$; $\ddot \theta \rightarrow \infty$.  The time step can always be decreased to surpress this issue (more than two orders of magnitude for this simulation), but for this research presentation the coordinate system for the spin moment dynamics was rotated about the $x$ axis by $\pi/2$.  Then once the simulation was complete the results were rotated back into the original coordinate system.

Initial spin orientation values, $\phi_i$ are divided equally from 0 to $\pi$ in increments of $\pi/1000$ (since the experiment starts well outside the Stern-Gerlach device, and inside the device the magnetic field is $\simeq B_z$, the $\phi_i$ angles are comparable to $\beta$ and will be compared as such).  Each $\phi_i$ value is given a random $\theta_i$ value and the simulation begins.  This action is repeated 1000 times for each $\phi_i$ with a new random $\theta_i$ resulting in a total of one million simulations.

\section{Carrier Kinematic Results}

The trajectory due to the carrier kinematics are broken into two main sections: one where the usual oversimplification of the force is used, as in equation \ref{eq:force_simp}, and where a full force is calculated, equation \ref{eq:force_approximation}.

\begin{figure}
\includegraphics[clip, trim={0cm 0cm 0cm 0cm}, scale=.55]{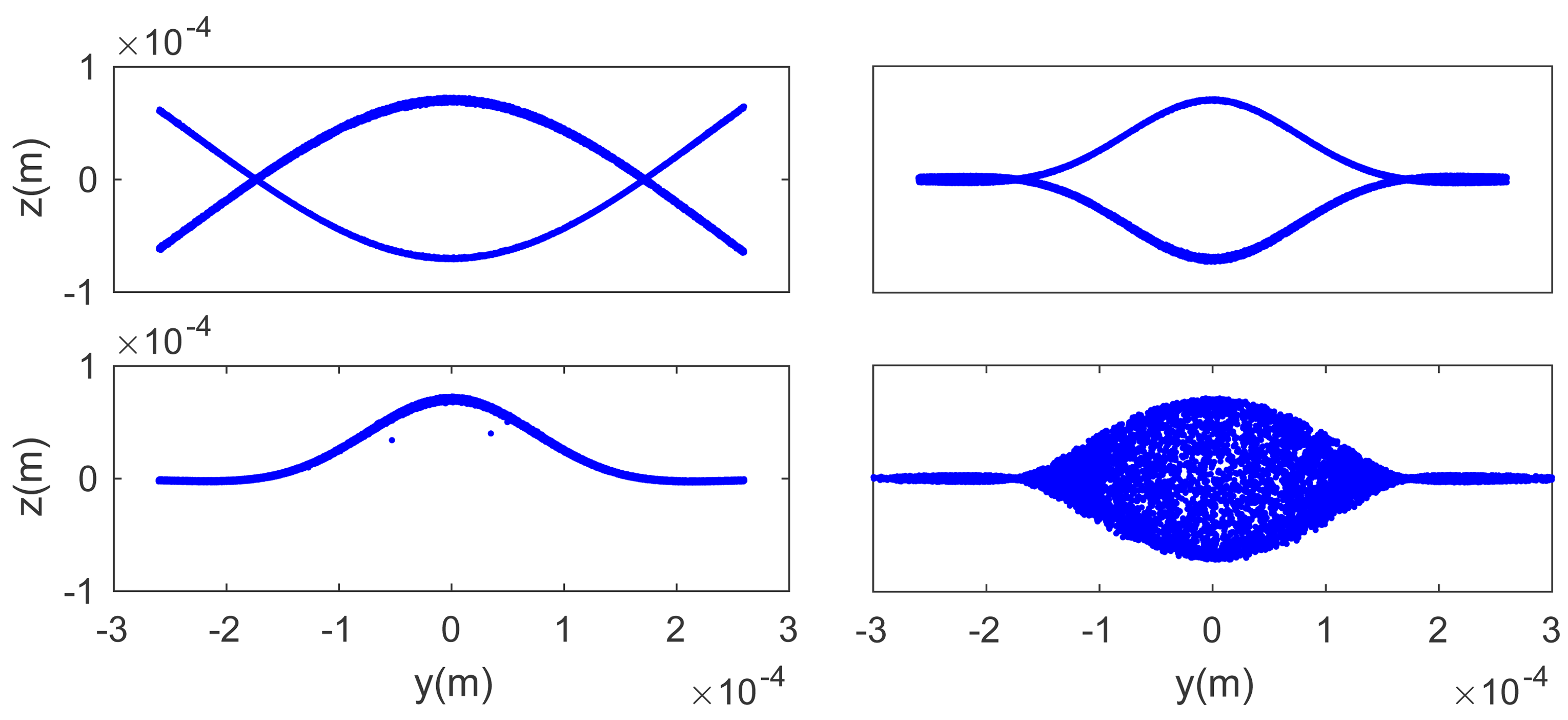}
\caption{\label{fig:trajectory}The classical trajectory split of the quantum spin due to equation \ref{eq:force_simp}.  The top left graph is the only trajectory that does not include the simplification in equation \ref{eq:divergence_approximation}.  The top right graph is an ideal trajectory from quantum spin.  The bottom left is using a classical dipole force shown in equation \ref{eq:torque_classical}.  The bottom right further assumes a large moment of inertia to a classical dipole.}
\end{figure}

Figure \ref{fig:trajectory} shows the results of the former.  It is interesting to note, that had the magnetic field been stronger, then the original Stern-Gerlach experiment would have had features similar to the top left graph where additional 'eyes' appear.  Once again that feature is due strictly from the field of the Stern-Gerlach device as seen in figure \ref{fig:divergence_B}.  The other trajectories take into account a minimum magnetic field needed to induce a quantum spin via equation \ref{eq:divergence_approximation}.

The top right graph is the most commonly perceived trajectory from the Stern-Gerlach Device.  The spins aligned with the magnetic field, spin up, experience a positive force as seen in figure \ref{fig:divergence_B}.  Had the experiment been rotated by $\pi$ around the $y$ axis and the magnetic field still pointing in the $\hat z$ direction, the gradient would be negative, $-\partial B / \partial z$, which would result in the spin up to go in the $-\hat z$ direction.  This concept is still misrepresented by many authors \cite{Liboff, Shankar, Wennerstrom} that a spin aligned with the magnetic field, spin up, will always have a lateral force towards the physical point-like structure in the Stern-Gerlach device.

The bottom trajectories in figure \ref{fig:trajectory} are implementing classical torque on a classical dipole moment.  The only difference is the plot to the right has two orders of magnitude higher moment of inertia than the plot on the left, which is the same as the quantum spin moment.

\begin{figure}
\includegraphics[clip, trim={5cm 11.5cm 5cm 11.5cm}, scale=.6]{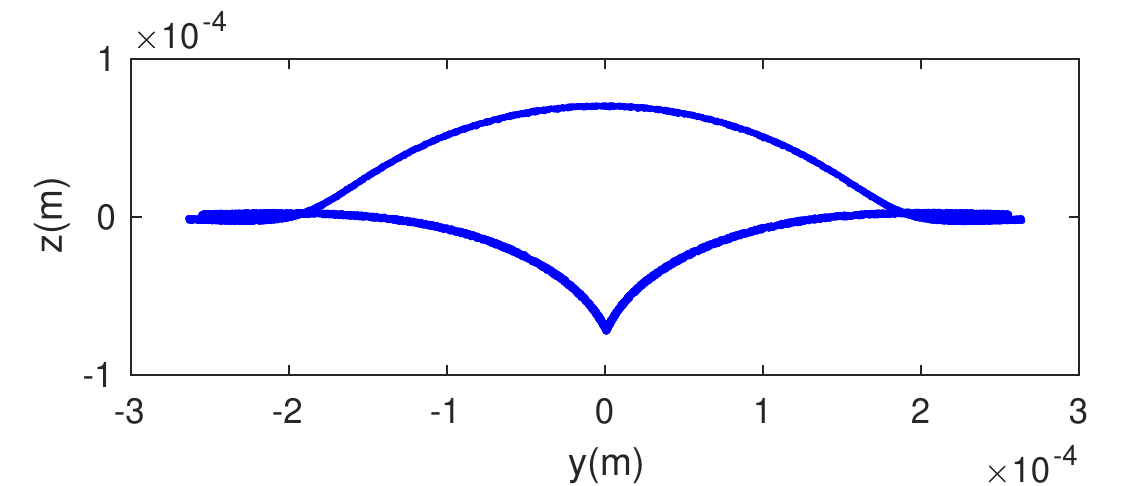}
\caption{\label{fig:trajectory_curl}The complete force classical trajectory split of the quantum spin, equation \ref{eq:force}.}
\end{figure}

\begin{figure}
\includegraphics[clip, trim={29cm 7cm 7.4cm 7cm}, scale=.4, angle =270]{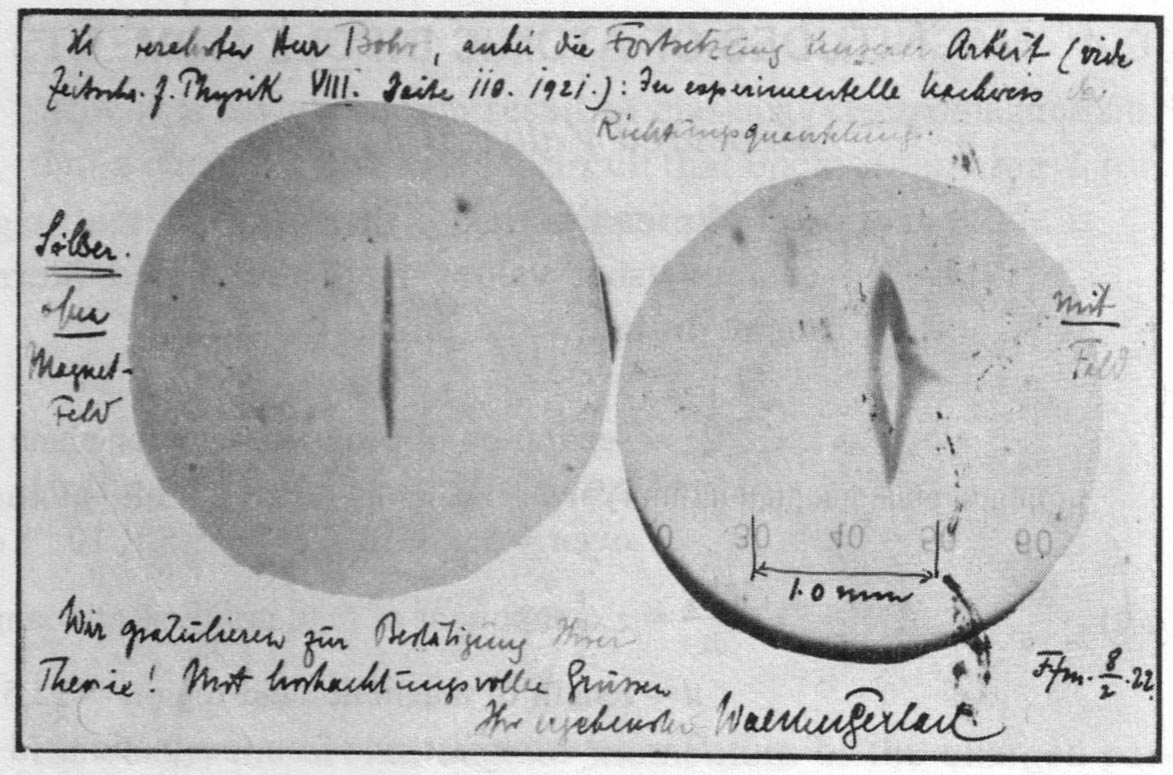}
\caption{\label{fig:letter_SGD}The original Stern-Gerlach results \cite{Stern} as seen in this postcard sent by Gerlach to Bohr \cite{SGD_letter}.}
\end{figure}

A full force calculation of the carrier kinematics is shown in figure \ref{fig:trajectory_curl}.  It is important to note that the drifting is caused due to the curling terms from analyzing the full force, equation \ref{eq:force}.  Spins aligned in the direction of the magnetic field, spin up, will always drift away from the center, $y=0$, whereas spin down particles will always drift towards the center regardless of flipping the magnetic field\footnote{The original Stern-Gerlach experiment \cite{Stern} has opposite drifting due to the structure of the bottom magnet.  The curling in a narrow region around $y=0$ is opposite and can been seen from magnetic field lines in a similar Stern-Gerlach-like device \cite{Gersem}}.

In comparing the full force calculation results of figure \ref{fig:trajectory_curl}, it looks very similar to the original results from the Stern-Gerlach experiment \cite{Stern}, as seen in figure \ref{fig:letter_SGD}.  In the original work, a depletion of spin up particles at $y=0$ is noticeable due to the drifting caused by the curling of the magnetic field.

\section{Moment Dynamic Results}

The process of the collapse of the individual quantum spin states needs to match known quantum statistics to be suggested as a possible model.  The statistics as shown in equation \ref{eq:Percentage} is the basis for comparision.

\begin{figure}
\includegraphics[clip, trim={0cm 0cm 0cm -1cm}, scale=.48]{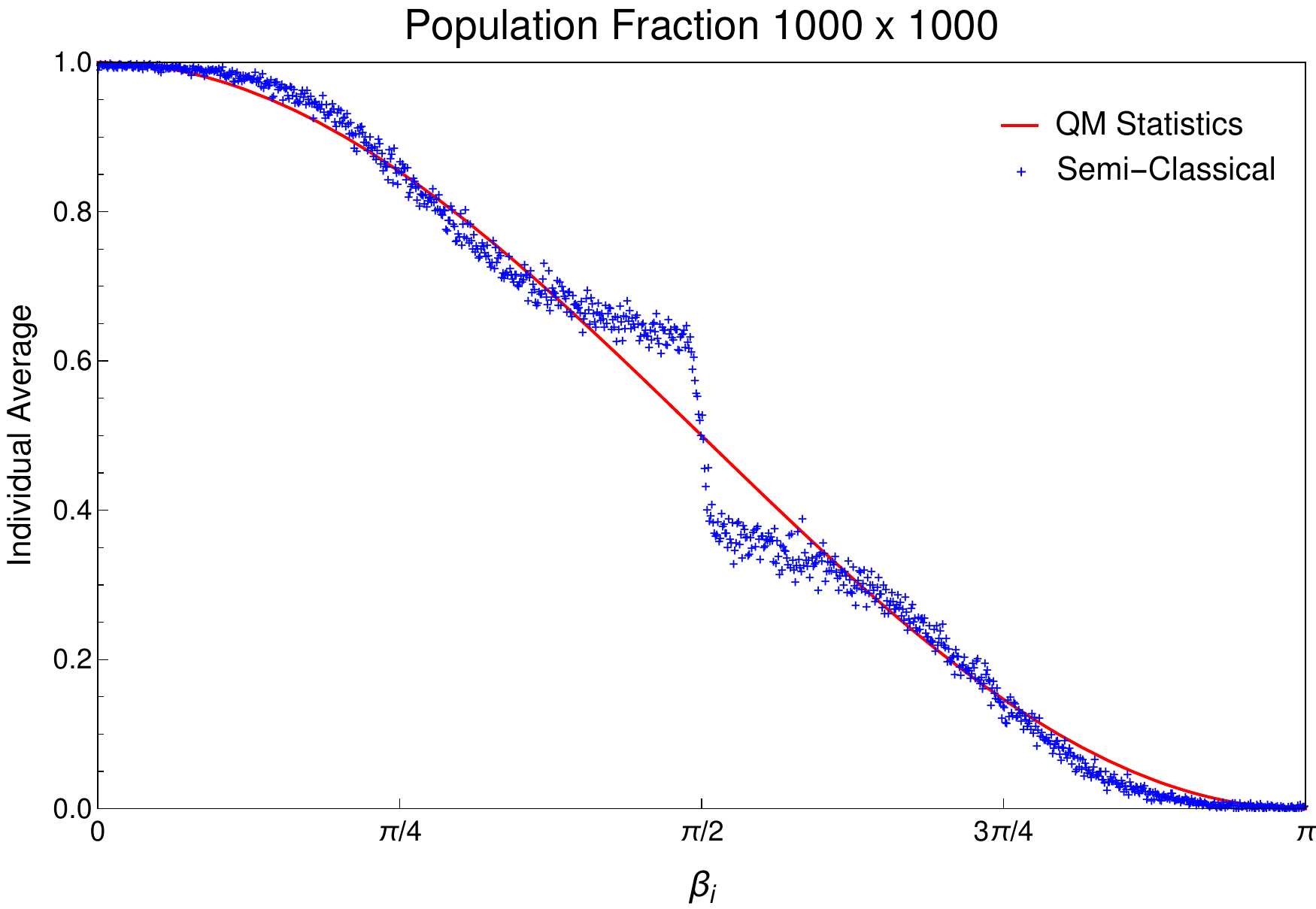}
\caption{\label{fig:QM_comparision}Comparision of the well known quantum spin statistics to the semi-classical model.  Each green marker is a fractional population of 1,000 individual runs.  For example a marker at $\pi/4$ represents 1,000 simulation results averaged.  That average represents the probability for a spin flip to occur.}
\end{figure}

The quantum mechanical statistic comparision to the model set forth in this research is shown in figure \ref{fig:QM_comparision}.  There were many adjustable parameters, but the best results are shown.

The piecewise-like behaviour around $\pi/2$ is particularly alarming since this should be a very unstable area and thus an equal opportunity for a spin to flip either up or down.

In further analysis of the trajectories, initial slices (i.e. restricting the initial azimuthal angle to certain values) were used to compare with the quantum spin statistics (since the probability should be irrelevant to azimuthal angle).  The discovery was that azimuthal angles close to 0 or $\pi$ had a really high value of flipping; whereas azimuthal angles around $\pi/2$ or $3\pi/4$ had relatively zero chance of flipping.

As stated in the original suggestion to model quantum mechanics using a nonlinear system that exhibits chaos, this system with just one perturbation like moment when $B_x/B_z>1$ is not enough to produce the chaos required to mimic quantum mechanical spin state probabilities.

\section{Conclusion}

Further insights were obtained into the carrier kinematics and therefore the trajectory results.  The cause of the translational force is due to the curling of the magnetic field, which is asymetric about the $x$-$z$ plane at $y=0$.

Although many attempts into changing the torque function and dampening parameter were done, the best results comparing the semi-classical quantum spin state results to known quantum statistics falls short of expectation (as seen in figure \ref{fig:QM_comparision}).

In this research the model sought chaos due to a perturbation perpendicular to the main magnetic field direction inside the Stern-Gerlach device.  Since there does not exist enough perturbation to cause the needed chaos, a different proposal is needed, perhaps internally in the spin model, to exhibit the highly chaotic behavior predicted in literature \cite{Thayer_Jafari}.

In spite of a full representation of the magnetic field given, the dynamics on a quantum spin due to opposing magnetic flux was not taken into account due to it's magnitude.  As stated earlier, research has yet to measure or predict the dynamics that must exist when a quantum spin is opposing the magnetic flux through it's finite size.  We encourage those with capabilities to show dynamics of a quantum spin due to opposing a magnetic flux to verify the universal flux law at a quantum level.

Furthermore, in looking towards the future at other requirements for the quantum spin, replication of Rabi oscillations is necessary.  To remind the reader, Rabi oscillation has a constant magnetic field and a perpendicular oscillating magnetic field:

\begin{equation}
\vec B = B_0 \hat z + B_1(\cos {\omega t} \, \hat x - \sin {\omega t} \, \hat y).
\end{equation}   

A full understanding of Rabi cycles shows that an oscilating field does not have to be larger than the dominant field i.e. $B_1/B_0\ngtr 1$, which does not bode well for the semi-classical model looking for chaos due to a perpendicular perturbation.  This further confirms the need to look elsewhere for chaos behaviour from the semi-classical quantum spin.

This work was supported by the National Science Foundation and the Royal Society of New Zealand under the East Asia and Pacific Summer Institutes Award Number: 1713790, the facilities at the Centre for Theoretical Physics and Chemistry at the Massey University Albany Campus along with the local HPC cluster, the University of Wyoming Physics and Astronomy department, the Mount Moran HPC cluster at the Advanced Research Computing Center \cite{Mtmoran}, and the Wyoming NASA Space Grant Consortium, NASA Grant \#NNX15AI08H.

\widetext
\clearpage
\begin{center}
\textbf{\large Supplementary Information}
\end{center}
\setcounter{equation}{0}
\setcounter{figure}{0}
\setcounter{table}{0}
\setcounter{page}{1}
\setcounter{section}{0}
\setcounter{table}{0}
\makeatletter
\renewcommand{\theequation}{S\arabic{equation}}
\renewcommand{\thefigure}{S\arabic{figure}}
\renewcommand{\thetable}{S\arabic{table}}
\renewcommand{\bibnumfmt}[1]{[S#1]}
\renewcommand{\citenumfont}[1]{S#1}

\section{Angular Acceleration in Spherical Coordinares} \label{sec:ang_acc}
Under spherical coordinates where $\phi$ is the angle off of the $z$ axis, $\theta$ is the angle around $z$ off of the $x$ axis as in figure \ref{fig:setup_figure} and letting $\vec r$ act as the rotational evolution of $\vec \mu$, the linear kinematics are,

\begin{equation}
\begin{split}
\vec r =& \, r \er, \\
\vec v =& \, \dot r \er + r\dot \theta \sp \et + r\dot \phi \ep \\
\vec a =& \left( \ddot r - r {\dot \phi}^2 - r{\dot \theta}^2 \sin^2 \phi\right) \hat r + \left(2 \dot r \dot \theta \sp + 2 r \dot \theta \dot \phi \cp + r \ddot \theta \sp\right) \hat \theta + \left( 2 \dot r \dot \phi + r \ddot \phi - r {\dot \theta}^2 \sp \cp \right) \hat \phi .
\end{split}
\end{equation}

\noindent The angular velocity is defined as,
\begin{equation}
\vec \omega = \frac{\vec r \times \vec v}{r^2},
\end{equation}

\noindent and differentiating this gives angular acceleration,

\begin{equation}
\vec \alpha = \frac{d \vec \omega}{dt} =
\cancelto{0}{\frac{\vec v \times \vec v}{r^2}} + \frac{\vec r \times \vec a}{r^2} - 2 \dot r \frac{\vec r \times \vec v}{r^3}.
\end{equation}

\noindent Taking cross products give,

\begin{align}
\begin{split}
\vec r \times \vec a &= \begin{vmatrix}
                        \hat r & \hat \theta & \hat \phi \\
                        r & 0 & 0 \\
                        a_r & a_\theta & a_\phi
                       \end{vmatrix} 
                      = -r a_\phi \et + r a_\theta \ep \\
                      &= -\left( 2 r\dot r \dot \phi + r^2 \ddot \phi - r^2 {\dot \theta}^2 \sp \cp \right) \hat \theta +
                      \left( 2 r\dot r \dot \theta \sp + 2 r^2 \dot \theta \dot \phi \cp + r^2 \ddot \theta \sp \right) \hat \phi,
\end{split}\\
\begin{split}
\vec r \times \vec v &= \begin{vmatrix}
                        \hat r & \hat \theta & \hat \phi \\
                        r & 0 & 0 \\
                        \dot r & r \dot \theta \sp & r \dot \phi
                       \end{vmatrix} \\
                     &= - r^2 \dot \phi \et + r^2 \dot \theta \sp \ep. 
\end{split}
\end{align}
Finally, put it all together to get,
\begin{equation}
\begin{split}
\vec \alpha =& -\left( 2 \frac{\dot r}{r} \dot \phi + \ddot \phi - {\dot \theta}^2 \sp \cp \right) \hat \theta +
\left( 2 \frac{\dot r}{r} \dot \theta \sp + 2 \dot \theta \dot \phi \cp + \ddot \theta \sp \right) \hat \phi
+ 2 \frac{\dot r}{r} \dot \phi \et - 2 \frac{\dot r}{r}  \dot \theta \sp \ep  \\
=& -\left( \ddot \phi - {\dot \theta}^2 \sp \cp \right) \hat \theta +
\left( 2 \dot \theta \dot \phi \cp + \ddot \theta \sp \right) \hat \phi,
\end{split}
\end{equation}
and rotate back to a Cartesian frame,
\begin{equation}
\begin{split}
\vec \alpha =& -\left( \ddot \phi - {\dot \theta}^2 \sp \cp \right) \left( \ct \ey - \st \ex \right) \\
&+ \left( 2 \dot \theta \dot \phi \cp + \ddot \theta \sp \right)  \left( \cp \ct \ex + \cp \st \ey - \sp \ez \right) \\
=& \begin{pmatrix}
    \ddot \phi \st - {\dot \theta}^2 \st \sp \cp + 2\dot \theta \dot \phi \ct \cos^2 \phi + 
    \ddot \theta \ct \sp \cp \\
    -\ddot \phi \ct + {\dot \theta}^2 \ct \sp \cp + 2\dot \theta \dot \phi \st \cos^2 \phi +
     \ddot \theta \st \sp \cp \\  
     -2\dot \theta \dot \phi \cp \sp - \ddot \theta \sin^2 \phi
    \end{pmatrix}.
\end{split}
\end{equation}
\section{Moment Dynamics: Equations of Motion} \label{sec:EOM}
In solving equation \ref{eq:EoM_torque}, but representing the torque due to a magnetic field in terms of a torque magnitude model that is a function of the angle between the magnetic moment and the magnetic field, TTM($\beta$), such that for any specific axis of rotation, $\hat k$, the torque will be represented as,

\begin{equation}
\vec \tau_k = \mu B \, \textrm{TMM}(\beta) \, \hat n \cdot \hat k,
\label{eq:torque_TMM}
\end{equation}

\noindent where,

\begin{equation}
\beta = \arccos\left( \frac{B_x \sp \ct + B_y \sp \st + B_z \cp}{\sqrt{B_x^2+B_y^2+B_z^2}}\right),
\end{equation}

\noindent and solving for $\ddot \phi$ and $\ddot \theta$,  

\begin{equation}
I 
\begin{bmatrix}
1 & 0 & 0 \\
0 & 1 & 0 \\
0 & 0 & 0
\end{bmatrix}
\begin{pmatrix}
    \ddot \phi \st - {\dot \theta}^2 \st \sp \cp + \\ 2\dot \theta \dot \phi \ct \cos^2 \phi + 
    \ddot \theta \ct \sp \cp \\[6pt]
    -\ddot \phi \ct + {\dot \theta}^2 \ct \sp \cp + \\ 2\dot \theta \dot \phi \st \cos^2 \phi +
     \ddot \theta \st \sp \cp \\[6pt]
     -2\dot \theta \dot \phi \cp \sp - \ddot \theta \sin^2 \phi
    \end{pmatrix}=
\begin{pmatrix}
    \mu B \, \textrm{TMM}(\beta) \, \hat n \cdot \hat x  -b(\dot \phi \st + \dot \theta \sp \cp \ct) \\[6pt]
    \mu B \, \textrm{TMM}(\beta) \, \hat n \cdot \hat y  -b(-\dot \phi \ct + \dot \theta \sp \cp \st) \\[6pt]
    \mu B \, \textrm{TMM}(\beta) \, \hat n \cdot \hat z  -b(-\dot \theta \sin^2 \phi)
    \end{pmatrix},
\end{equation}

\noindent due to the moment of inertia tensor, the problem can be simplified into a system of two linear equations,

\begin{equation}
\begin{cases}
\begin{array}{ccc}
\ddot \phi \st + \ddot \theta \ct \sp \cp & = & 
                \begin{matrix}I^{-1}\big(\mu B \, \textrm{TMM}(\beta) \, \hat n \cdot \hat x  -b(\dot \phi \st + \dot \theta \sp \cp \ct)\\ + {\dot \theta}^2 \st \sp \cp - 2\dot \theta \dot \phi \ct \cos^2 \phi\big)
                \end{matrix}\\[11pt]
-\ddot \phi \ct + \ddot \theta \st \sp \cp & = & 
                \begin{matrix}I^{-1}\big(\mu B \, \textrm{TMM}(\beta) \, \hat n \cdot \hat y  -b(-\dot \phi \ct + \dot \theta \sp \cp \st) \\- {\dot \theta}^2 \ct \sp \cp - 2\dot \theta \dot \phi \st \cos^2 \phi\big)
                \end{matrix}\\
\end{array}
\end{cases},
\end{equation}

\noindent which in linear algebra form, $AX=B$, and letting $m$ and $n$ equal the values on the right hand side, is:

\begin{equation}
\begin{bmatrix}
\st & \ct \sp \cp \\
-\ct & \st \sp \cp \\
\end{bmatrix}
\begin{bmatrix}
\ddot \phi \\
\ddot \theta \\
\end{bmatrix}
= 
\begin{bmatrix}
m\\n\\
\end{bmatrix}.
\end{equation}

\noindent Where the inverse of our $A$ matrix is,

\begin{equation}
\begin{bmatrix}
\st & \ct \sp \cp \\
-\ct & \st \sp \cp \\
\end{bmatrix}^{-1}
=
\begin{bmatrix}
\st & -\ct \\
\ct \csc \phi \sec \phi & \st \csc \phi \sec \phi \\
\end{bmatrix}.
\end{equation}

\noindent Multiplying $A^{-1}$ to both sides, the equations of motion are,

\begin{equation}
\begin{bmatrix}
\ddot \phi \\
\ddot \theta \\
\end{bmatrix}
= 
\begin{bmatrix}
m \st - n \ct \\
\csc\phi\sec\phi(m\ct + n \st)\\
\end{bmatrix}.
\end{equation}

Finally, a full equation is obtained by complete substitution and carrying out the dot product between the unit vector of magnetic torque rotation and the respective unit vectors.  Furthermore, the spin moment vector components are represented by $\phi$ and $\theta$ in cartesian coordinates as seen in figure \ref{fig:setup_figure},

\begin{equation} \label{eq:EOM_full}
\begin{bmatrix}
\ddot \phi \\
\ddot \theta \\
\end{bmatrix}
= 
\begin{bmatrix}
I^{-1}\big[\st\Big(\mu B \, \textrm{TMM}(\beta) \, \left(B_z \st \sp - B_y \cp\right)\big((B_x \cp - B_z \ct \sp)^2 \\
 + (B_y \ct \sp - B_x \st \sp)^2 + (B_z \st \sp - B_y \cp)^2\big)^{-1/2}\\
-b(\dot \phi \st + \dot \theta \sp \cp \ct) + {\dot \theta}^2 \st \sp \cp - 2\dot \theta \dot \phi \ct \cos^2 \phi\Big)\\
 - \ct\Big(\mu B \, \textrm{TMM}(\beta) \, \left(B_x \cp - B_z \ct \sp\right)\\
\big((B_x \cp - B_z \ct \sp)^2 + (B_y \ct \sp - B_x \st \sp)^2 + (B_z \st \sp - B_y \cp)^2\big)^{-1/2}\\
 -b(-\dot \phi \ct + \dot \theta \sp \cp \st) - {\dot \theta}^2 \ct \sp \cp - 2\dot \theta \dot \phi \st \cos^2 \phi\Big)\big] \\[8pt]
I^{-1}\csc\phi\sec\phi\bigg(\ct\Big(\mu B \, \textrm{TMM}(\beta) \, \left(B_z \st \sp - B_y \cp\right)\big((B_x \cp - B_z \ct \sp)^2\\
 + (B_y \ct \sp - B_x \st \sp)^2 + (B_z \st \sp - B_y \cp)^2\big)^{-1/2}\\
-b(\dot \phi \st + \dot \theta \sp \cp \ct) + {\dot \theta}^2 \st \sp \cp - 2\dot \theta \dot \phi \ct \cos^2 \phi\Big)\\ 
+ \st \Big(\mu B \, \textrm{TMM}(\beta) \, \left(B_x \cp - B_z \ct \sp\right)\\
\big((B_x \cp - B_z \ct \sp)^2 + (B_y \ct \sp - B_x \st \sp)^2 + (B_z \st \sp - B_y \cp)^2\big)^{-1/2}\\
 -b(-\dot \phi \ct + \dot \theta \sp \cp \st) - {\dot \theta}^2 \ct \sp \cp - 2\dot \theta \dot \phi \st \cos^2 \phi\Big) \bigg)\\
\end{bmatrix}.
\end{equation}

Now one can appreciate leaving the torque as TMM($\beta$) since this allows the user to quickly change out different torque models and even substitute in the classical torque model.  Therefore this equation of motion is good for classical and semi-classical depending on the torque model used (this versatility can be seen in figure \ref{fig:trajectory}).

\section{Stern-Gerlach Device Dimensions} \label{sec:SGD}

\begin{figure}[!htb]
\includegraphics[clip, trim={0cm 0cm 0cm 0cm}, scale=.7]{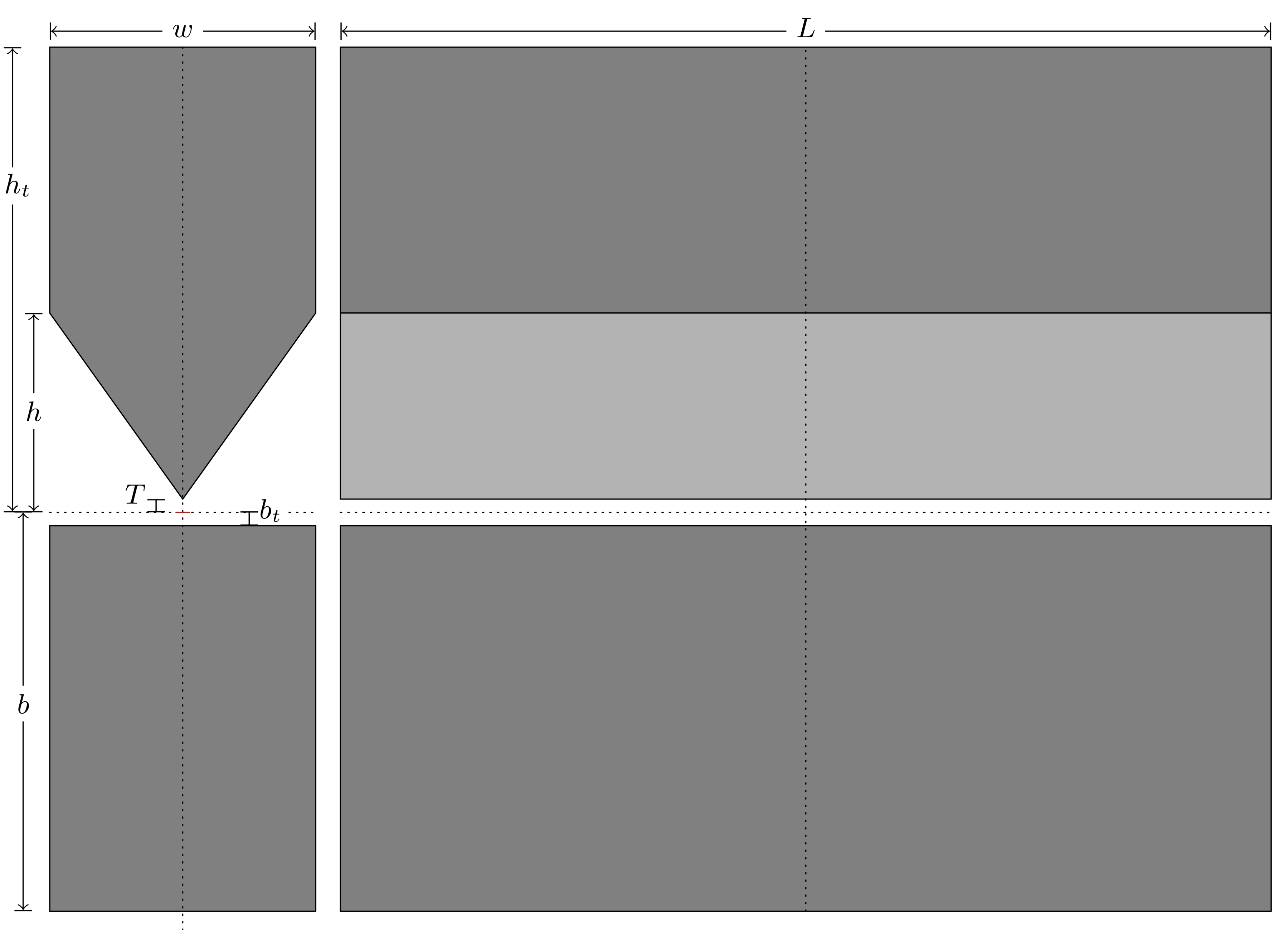}
\caption{The dimensions and definitions of the Stern-Gerlach device used in this research.  The red line shows the incoming beam width into the device.}
\label{fig:setup_both}
\end{figure}

The magnetic field for the Stern-Gerlach device.  The tip angle used was based off the original research \cite{Sref:Stern}.  The overall view, seen in figure \ref{fig:setup_figure}, is shown in terms of figurative dimensions in figure \ref{fig:setup_both} and following definitions in table \ref{tab:SGD}.

\begin{table}[h!]
\centering
\begin{tabular}{|c|c|c|e{2.2}|}
\hline
&&&\\[-.8em]
Symbol & Meaning & Definition & \mc{Value (\si{\cm}) } \\
&&&\\[-.8em]
\hline
\\[-.85em]
$w$ & Width & Width of both the top and bottom pieces & 1.0{ }\\
$h_t$ & Top Height & Height from the center to the top piece & 1.75\\
$b$ & Bottom Height & Height from the bottom piece to the center & 1.5\\
$h$ & Tip Length & Height from the center to the beginning of the top tip & .75\\
$T$ & Tip & Distance from tip point to center & .05\\
$b_t$ & Bottom Top & Distance from the top of the bottom piece to the center & .05\\
$L$ & Length & Overall length of the Stern-Gerlach device & 3.5\\[.15em]
\hline
\end{tabular}
\caption{Further description of the symbols used in figure \ref{fig:setup_both}.}
\label{tab:SGD}
\end{table}

The simulation begins 1\si{\cm} away from the Stern-Gerlach device as shown in figure \ref{fig:setup_figure}.  The initial random starting location on the $y$-$z$ plane was a square centered around the $x$ axis was confined between $\pm$\SI{1}{\micro\metre} $\hat y$ and $\pm$\SI{1}{\micro\metre} $\hat z$.  The magnetic field strength of the Stern-Gerlach device was assumed to be the saturation value of iron (although in literature it was aided by a wire carrying current wrapped around the iron).

\end{document}